\documentclass[aps,twocolumn,showpacs,preprintnumbers,amsmath,amssymb,superscriptaddress,floatfix,nofootinbib]{revtex4-1}
\usepackage[left=2cm,right=2cm]{geometry}
\usepackage{graphicx}
\usepackage{epsfig,float}

\usepackage{hyperref}
\usepackage{amsmath}
\usepackage{amsfonts}
\usepackage{amssymb}

\begin{document}

\title{Predictions for the $\bar B^0 \to \bar K^{*0} X (YZ)$ and  $\bar B^0_s \to \phi X (YZ)$ with  $X(4160), Y(3940), Z(3930)$}

\author{Wei-Hong Liang}
%\email{liangwh@gxnu.edu.cn}
\affiliation{Department of Physics, Guangxi Normal University,
Guilin 541004, China}
\affiliation{Institute of Modern Physics, Chinese Academy of
Sciences, Lanzhou 730000, China}

\author{R. Molina}
%\email{liangwh@gxnu.edu.cn}
\affiliation{The George Washington University, 725 21st St NW, Washington, DC 20052,
USA}

\author{Ju-Jun Xie}

\affiliation{Institute of Modern Physics, Chinese Academy of
Sciences, Lanzhou 730000, China} \affiliation{Research Center for
Hadron and CSR Physics, Institute of Modern Physics of CAS and
Lanzhou University, Lanzhou 730000, China} \affiliation{State Key
Laboratory of Theoretical Physics, Institute of Theoretical Physics,
Chinese Academy of Sciences, Beijing 100190, China}

\author{M.  D\"oring}
%\email{liangwh@gxnu.edu.cn}
\affiliation{The George Washington University, 725 21st St NW, Washington, DC 20052,
USA}

\author{E.~Oset}

\affiliation{Institute of Modern Physics, Chinese Academy of
Sciences, Lanzhou 730000, China} \affiliation{Departamento de
F\'{\i}sica Te\'orica and IFIC, Centro Mixto Universidad de
Valencia-CSIC Institutos de Investigaci\'on de Paterna, Aptdo.
22085, 46071 Valencia, Spain}

\date{\today}

\begin{abstract}

We investigate the decay of $\bar B^0 \to \bar K^{*0} R$ and  $\bar B^0_s \to \phi R$ with  $R$ being the $X(4160)$, $Y(3940)$, $Z(3930)$ resonances.  Under the assumption that these states are dynamically generated from the vector-vector interaction, as has been concluded from several theoretical studies, we use a reaction mechanism of quark production at the elementary level, followed by hadronization of one final $q \bar q$ pair into two vectors and posterior final state interaction of this pair of vector mesons to produce the resonances. With this procedure we are able to predict five ratios for these decays, which are closely linked to the dynamical nature of these states, and also predict the order of magnitude of the branching ratios which we find of the order of $10^{-4}$, well within the present measurable range.
In order to further test the dynamical nature of these resonances we study the $\bar B^0_s \to \phi D^* \bar D^*$ and $\bar B^0_s \to \phi D_s^* \bar D_s^*$ decays close to the $D^* \bar D^*$ and $D_s^* \bar D_s^*$ thresholds and make predictions for the ratio of the mass distributions in these decays and the $\bar B^0_s \to \phi R$ decay widths. The measurement of these decays rates can help unravel the nature of these resonances.

\end{abstract}

\maketitle

\section{Introduction}

The $XYZ$ resonances with masses in the region around 4000 MeV have stirred the hadron community with a series of states that challenge the common wisdom of mesons as made from $q \bar q$. There has been intense experimental work done at the BABAR, BELLE, CLEO, BES and other collaborations, and many hopes are pinned in the role that the future FAIR facility with the PANDA collaboration and J-PARC will play in this
field.  There are early experimental reviews on the topic  \cite{Barnes:2006xq,Asner:2008nq,Godfrey:2008nc,ChengPing:2009vu} and more recent ones \cite{Olsen:2012zz,Lange:2013sxa,Liu:2013waa,Guo:2014pya,Olsen:2014qna}. From the theoretical point of view there has also been an intensive activity trying to understand these intriguing states. There are quark model pictures \cite{Ortega:2012rs,Vijande:2014cfa} and explicit tetraquark structures \cite{Esposito:2014rxa}. Molecular interpretations are given in refs. \cite{Branz:2009yt,Branz:2010sh,Yang:2009fj,Dong:2012hc,Wang:2013cya,Cleven:2013sq,Wang:2013kra,
Aceti:2014kja,Aceti:2014uea}. Much progress also has been done using the heavy quark spin symmetry (HQSS) \cite{Nieves:2012tt,HidalgoDuque:2012pq}. Predictions using QCD sum rules have also brought some light into the issue \cite{Nielsen:2009uh,Khemchandani:2013iwa,Nielsen:2014mva}. Strong decays of these resonances have been studied to learn about the nature of these states \cite{Dong:2013iqa,Ma:2014zva}, while very often radiative decays are invoked as a tool to provide insight into this problem \cite{Liang:2009sp,Branz:2010gd,Aceti:2012cb,Ma:2014ofa,Dong:2014zka}, although there might be exceptions as discussed in ref. \cite{Guo:2014taa}. It has even been speculated that some states found near thresholds of two mesons could just be cusps, or threshold effects \cite{Swanson:2014tra}.
However, this speculation was challenged in ref. \cite{Guo:2014iya} which showed that
the near threshold narrow structures cannot be simply explained by
kinematical threshold cusps in the corresponding elastic channels but
require the presence of $S$-matrix poles.
Along this latter point one should also mention a recent work calling
the attention to possible effects of singularities on the opposite side
of the unitary cut that enhance the cusp structure for states with mass
above a threshold \cite{Szczepaniak:2015eza}.
 Some theoretical reports on these issues can be found in
refs. \cite{Zhu:2007wz,Brambilla:2010cs,Drenska:2010kg}.

  On the other hand, and somewhat unexpected, recent experiments on the weak decays of $B$ meson are proving to be a powerful source of information on hadron dynamics and the nature of hadronic states \cite{Aaij:2011fx,Li:2011pg,Aaltonen:2011nk,Abazov:2011hv,Aaij:2013zpt,Aaij:2014siy}. One of the recent surprises was to see from these experiments a pronounced peak for the $f_0(980)$ in $B^0_s$ decay into $J/\psi$ and $\pi^+ \pi^-$ \cite{Aaij:2011fx} while the signal for the $f_0(500)$ was found very small or non-existent. Simultaneously, in the analogous decay of $\bar B^0$ into $J/\psi$ and $\pi^+ \pi^-$ \cite{Aaij:2013zpt} a signal was seen for the $f_0(500)$ production and only a very small fraction was observed for the $f_0(980)$ production. This is surprising since the $f_0(500)$ couples more strongly to $\pi^+ \pi^-$ than to the $f_0(980)$. Some attempt to explain this behaviour was offered in ref. \cite{stone} in terms of tetraquark structures for the scalar mesons. Furthermore, the $f_0(500)$ and $f_0(980)$ resonances are naturally explained as states dynamically generated from the meson-meson interaction in chiral unitary theory \cite{npa,ramonet,kaiser,markushin,juanito,rios}.  From this perspective the features observed in these experiments and different ratios were well described in ref. \cite{weihong}. The basic approach in ref. \cite{weihong} was to identify the dominant mechanism of the decays at the quark level, implementing the hadronization of the final $q \bar q$ pair into two mesons and allowing them to interact to generate the $f_0(500)$ and $f_0(980)$ resonances.

  The picture of ref. \cite{weihong} has been extended to describe many other $B$ and $D$  decays, with equal success in the description of observed features of the reactions. In ref. \cite{Bayar:2014qha} ratios for the production of vector mesons in the final states were evaluated and predictions for the $\bar B^0_s \to J/\psi \kappa(800)$ decay were made. In ref. \cite{daid} the  $D^0$ decays into $K^0_s$ and $f_0(500)$, $f_0(980)$, $a_0(980)$ were studied. A different sort of states, dynamically generated from the vector-vector interaction were investigated in ref. \cite{xievec} in the $\bar{B}^0$ and $\bar{B}^0_s$ decays into $J/\psi$ and $f_0(1370),~f_0(1710),~f_2(1270),~f'_2(1525),~K^*_2(1430)$. In ref. \cite{xiebd} the $\bar B^0$ decay into $D^0$ and $\rho$ or $f_0(500)$, $f_0(980)$, $a_0(980)$ and $\bar B^0_s$ decay into $D^0$ and $K^{*0}$ or $\kappa(800)$ were studied. Similarly, the  $B^0_s$ decay into $D_s~DK$ was studied with the aim of learning about $KD$ scattering and the $D_{s0}^*(2317)$ resonance \cite{marina}. Related work to this latter one was done studying the semileptonic $B_s$ and $B$ decays in ref. \cite{fernando}.

Related problems of these weak decays have been addressed in refs. \cite{robert,bruno,cheng,bruno2,lucio,bruno3,loiseau} from a very different point of view, evaluating microscopically the weak matrix elements and parametrizing parts of this interaction plus properties of the resonances produced, then  carrying fits to data. The aims are also usually different, having in mind the construction of full amplitudes that can be used for issues like $CP$ violation. On the contrary, the work of ref. \cite{weihong} and the related ones make predictions for shapes and ratios that are tied to the final state interaction of mesons and, consequently, parameter free predictions can be made for these observables.

 So far, in the study of these $B$ decays the production of $XYZ$ states has not yet been addressed and the aim of the present paper is to study reactions where these states can be produced, evaluating ratios for different decay modes and estimating the absolute rates. This should stimulate experimental work that can shed light on the nature of some of these controversial states.

\section{Formalism}
\begin{figure}[ht]
\begin{center}
\includegraphics[scale=0.55]{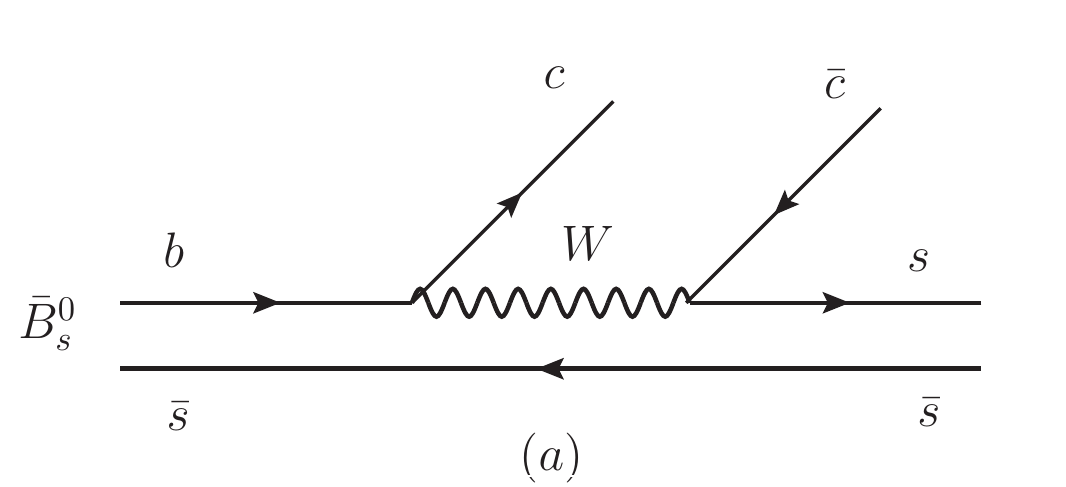}
\includegraphics[scale=0.55]{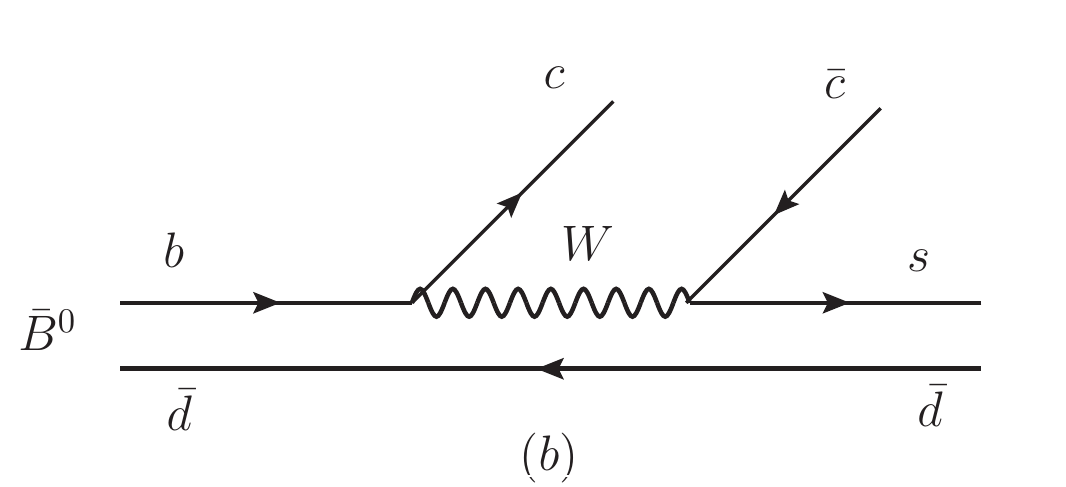}
\end{center}
\caption{Diagrams at the quark level for $\bar B^0_s$ (a) and $\bar B^0$ (b) decays into $c\bar c$ and a $q\bar q$ pair.}
\label{Fig:1}
\end{figure}

Following refs. \cite{stone, weihong}, we plot in Fig. \ref{Fig:1} the basic mechanism at the quark level for $\bar B^0_s (\bar B^0)$ decay into a final $c \bar c$ and another $q\bar q$ pair.
In ref. \cite{weihong} the $c\bar c$ went into the production of a $J/\psi$ and the $s\bar s$ or $s\bar d$ were hadronized to produce two mesons which were allowed to interact to produce some resonant states. Here, we shall follow a different strategy and allow the $c\bar c$ to hadronize into two vector mesons, while the $s\bar s$ and $s\bar d$ will make the $\phi$ and $\bar K^{*0}$ mesons respectively. Let us observe that, apart for the $b \to c$ transition, most favored for the decay, we have selected an $s$ in the final state which makes the $c\to s$ transition Cabibbo allowed. This choice magnifies the decay rate, which should then be of the same order of magnitude as the $\bar B_s^0 \to J/\psi f_0(980)$, which also had the same diagram of the quark level prior to the hadronization of the $s\bar s$ to produce two mesons, in this case $K\bar K$ that couples later to the $f_0(980)$.

\begin{figure*}[ht]
\begin{center}
\includegraphics[scale=0.55]{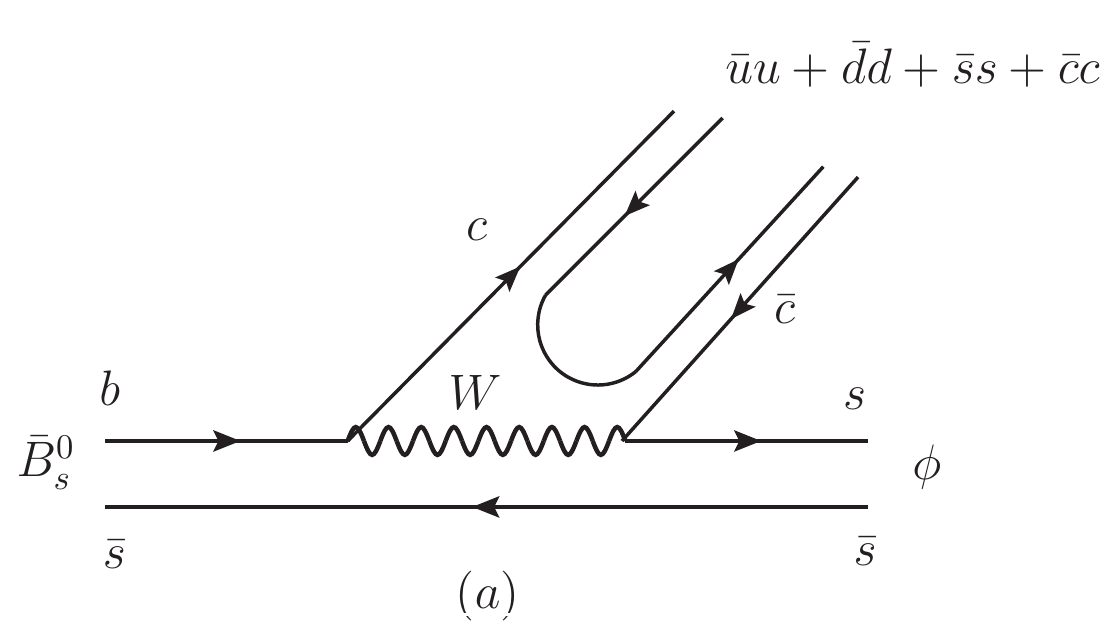}
\includegraphics[scale=0.55]{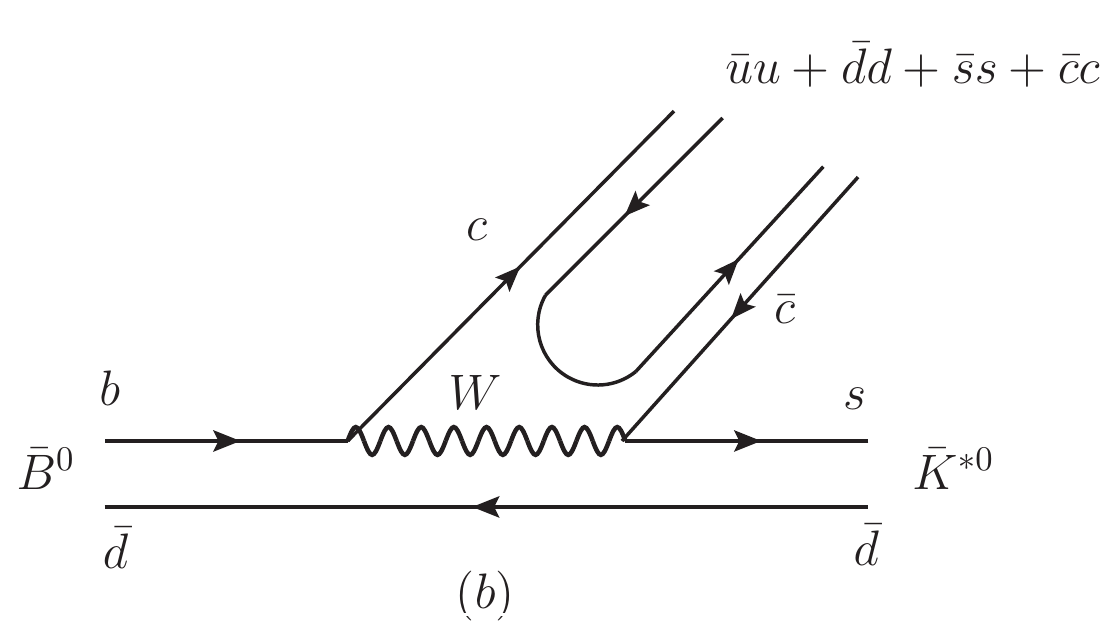}
\end{center}
\caption{Hadronization of the $c\bar c$ pair into two vector mesons for $\bar B^0_s$ decay (a) and $\bar B^0$ decay (b).} \label{Fig:2}
\end{figure*}

The next step consists in introducing a new $q\bar q$ state with the quantum numbers of the vacuum, $\bar uu +\bar d d +\bar s s +\bar c c$, and see which combinations of mesons appear when added to $c\bar c$. This is depicted in Fig. \ref{Fig:2}.
An easy way to see which vector mesons are produced in the hadronization of $c\bar c$ is to introduce the $q\bar q$ matrix
\begin{eqnarray}
M = \left(
           \begin{array}{cccc}
             u\bar u & u \bar d & u\bar s & u\bar c\\
             d\bar u & d\bar d & d\bar s & d\bar c\\
             s\bar u & s\bar d & s\bar s & s\bar c\\
             c\bar u & c\bar d & c\bar s & c\bar c\\
           \end{array}
         \right) = \left(
           \begin{array}{c}
            u   \\
             d  \\
             s   \\
             c   \\
           \end{array}
         \right) \left(
           \begin{array}{cccc}
            \bar{u} & \bar{d} & \bar{s} & \bar{c}
           \end{array}   \right).
\end{eqnarray}
Note that this matrix corresponds to the SU(4) vector matrix
\begin{equation}\label{eq:Vmatrix}
\renewcommand{\arraystretch}{1.5}
V = \left(
           \begin{array}{cccc}
             \frac{\rho^0}{\sqrt{2}} + \frac{\omega}{\sqrt{2}} & \rho^+ & K^{*+} & \bar D^{*0}\\
             \rho^- & -\frac{\rho^0}{\sqrt{2}} + \frac{\omega}{\sqrt{2}}  & K^{*0} & \bar D^{*-}\\
            K^{*-} & \bar{K}^{*0} & \phi & \bar D^{*-}_s\\
            D^{*0} & D^{*+} & D^{*+}_s & J/\psi\\
           \end{array}
         \right) .
\end{equation}
Now we see that \cite{alberzou}
\begin{eqnarray}
M \cdot M &=& \left(
           \begin{array}{c}
            u   \\
             d  \\
             s   \\
             c \\
           \end{array}
         \right) \left(
           \begin{array}{cccc}
            \bar{u} & \bar{d} & \bar{s} & \bar{c}
           \end{array}   \right)
           \left(
           \begin{array}{c}
            u   \\
             d  \\
             s   \\
             c \\
           \end{array}
         \right) \left(
           \begin{array}{cccc}
            \bar{u} & \bar{d} & \bar{s} & \bar{c}
           \end{array}   \right) \nonumber \\
           &=& \left(
           \begin{array}{c}
            u   \\
             d  \\
             s   \\
             c  \\
           \end{array}
         \right) \left(
           \begin{array}{cccc}
            \bar{u} & \bar{d} & \bar{s}   & \bar{c}
           \end{array}   \right) (\bar{u}u + \bar{d}d + \bar{s}s+ \bar{c}c)
           \nonumber \\
           &=& M (\bar{u}u + \bar{d}d + \bar{s}s + \bar{c}c). \label{MdotM}
\end{eqnarray}
Hence, we can write
\begin{equation}\label{eq:MM44}
c\bar c (\bar uu +\bar d d +\bar s s + \bar c c)\equiv (M \cdot M)_{44} \equiv (V \cdot V)_{44}
\end{equation}
and
\begin{equation}\label{eq:VV44-1}
(V \cdot V)_{44}= D^{*0} \bar D^{*0} + D^{*+} D^{*-} +D^{*+}_s  D^{*-}_s + J/\psi J/\psi.
\end{equation}
Note that we have produced an $I=0$ combination, as it should be coming from $c\bar c$ and the strong interaction hadronization, since we have the isospin doublet ($D^{*+}, -D^{*0}$), ($\bar D^{*0}, \bar D^{*-}$).
The $J/\psi J/\psi$ component is energetically forbidden and hence we can write
\begin{equation}\label{eq:VV44-2}
(V \cdot V)_{44} \to \sqrt{2} ( D^* \bar D^*)^{I=0} + D^{*+}_s  D^{*-}_s.
\end{equation}

We can reach the result of Eq. (\ref{eq:VV44-1}) in a simpler way neglecting the $c\bar c c \bar c$ component in Eq. (\ref{eq:MM44}). Since $c \bar c$ is an SU(3) singlet and so is the $\bar uu +\bar d d +\bar s s$  combination, then the $D^* \bar D^*$ combination of Eq. (\ref{eq:VV44-2}) is also an SU(3) singlet. Using the full formalism has the advantage that it already tells one the phase convention of the states, which should be taken consistently with the chiral study of the interaction based on the matrix of Eq. (\ref{eq:Vmatrix}).

Following the philosophy of ref. \cite{weihong} we shall now let these vector mesons undergo interaction and we can connect with the work of ref. \cite{raquelxyz}, where using an extension of the local hidden gauge approach \cite{hidden1,hidden2,hidden3,hidden4} some $XYZ$ states were dynamically generated. Concretely, in ref. \cite{raquelxyz} four resonances were found, that we summarize in Table \ref{Tab:XYZstate} together with the channel to which the resonance couples most strongly, and the experimental state to which they were associated.
\begin{table}[ht]
     \renewcommand{\arraystretch}{1.2}
\centering
\begin{tabular}{cccccc}
\hline\hline
Energy [MeV]~   & $I^G [J^{PC}]$ & Strongest & Experimental\\
& & channel &state\\
\hline
$3943-i7.4$   & ~$0^+ [0^{+~+}]$~ & $D^* \bar D^*$ & $Y(3940)$ \cite{Abe:2004zs}\\
$3945-i0$  & $0^- [1^{+~-}]$ & $D^* \bar D^*$ & ?~$Y_P$ \\
$3922-i26$  & $0^+ [2^{+~+}]$ & $D^* \bar D^*$  & $Z(3930)$ \cite{Uehara:2005qd}\\
$4169-i66$  & $0^+ [2^{+~+}]$ & $D^*_s \bar D^*_s$& $X(4160)$ \cite{Abe:2007sya}\\
\hline\hline
\end{tabular}
\caption{States found in  ref. \cite{raquelxyz}, the channel to which they couple most strongly, and the experimental states to which they are associated (see also refs. \cite{olsenrev, pdg}). $Y_P$ is a predicted resonance.}
\label{Tab:XYZstate}
\end{table}
In  ref. \cite{raquelxyz}, another state with $I=1$ was found, but this one cannot be produced with the hadronization of $c\bar c$. Some of these resonances have also been claimed to be of $D^* \bar D^*$ or $D^*_s \bar D^*_s$ molecular nature in refs. \cite{thomas,Gutsche:2010jf,valery} using for it the Weinberg compositeness condition \cite{weinberg,baru,hyodo} and in refs. \cite{saopaulo,wang,marinareview} using QCD sum rules. Also using HQSS the same conclusions are reached in ref. \cite{carlos} and with phenomenological potentials in ref. \cite{xiangliu}.

\begin{figure*}[ht]
\begin{center}
\includegraphics[scale=0.55]{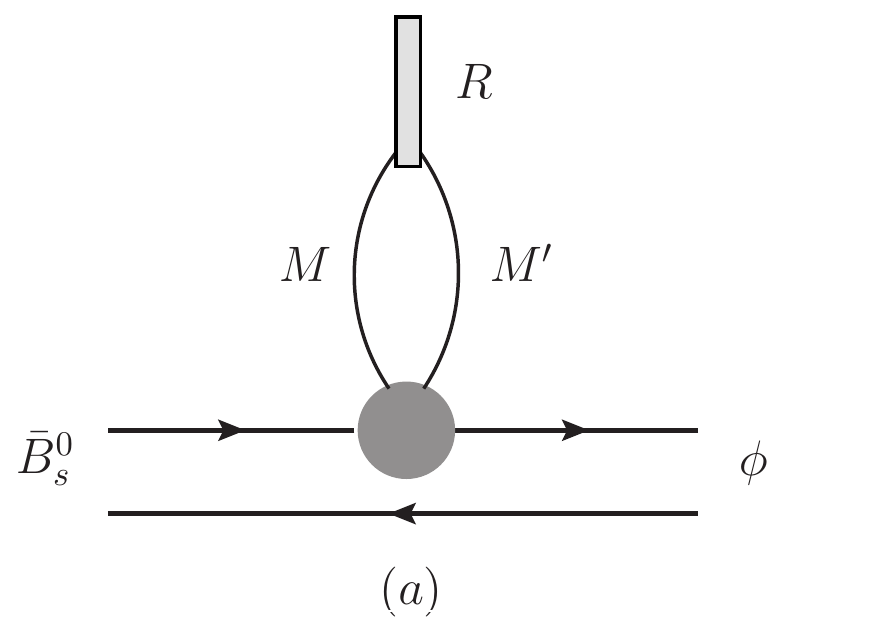}
\includegraphics[scale=0.55]{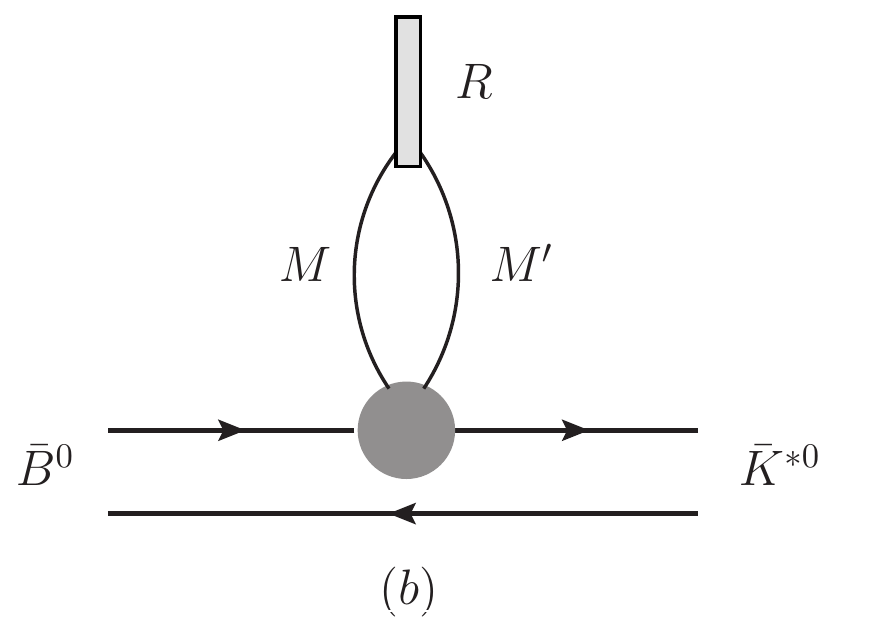}
\end{center}
\caption{Diagrammatic representation of the formation of the resonances $R (X,Y,Z)$ through rescattering of $M M'$ ($D^* \bar D^*$ or $D^*_s \bar D^*_s$) and coupling to the resonance.} \label{Fig:3}
\end{figure*}

The final state interaction of the $D^* \bar D^*$ and $D^*_s \bar D^*_s$ proceeds diagrammatically as depicted in Fig. \ref{Fig:3}.
Starting from Eq. (\ref{eq:VV44-2}) the analytical expression for the formation of the resonance $R$ is given by
\begin{equation}\label{eq:amplituT}
t(\bar B^0_s \to\phi R) = V_P (\sqrt{2} g_{D^* \bar D^*, R} G_{D^* \bar D^*}  + g_{D^*_s \bar D^*_s, R} G_{D^*_s \bar D^*_s}),
\end{equation}
where $G_{MM'}$ is the loop function of the two intermediate meson propagators and $g_{MM', R}$ is the coupling of the resonance to the $MM'$ meson pair.

The formalism for $\bar B^0 \to \bar K^{*0} R$ runs parallel since the hadronization procedure is identical, coming from the $c\bar c$, only the final state of $q\bar q$ is the $\bar K^{*0}$ rather than the $\phi$. Hence, the matrix element is identical to the one of $\bar B^0_s \to \phi R$, only the kinematics to different masses will change.

There is one last point to consider which is the angular momentum conservation. For $J^P_R =0^+,2^+$, we have the transition $0^- \to J^P~1^-$. Parity is not conserved but the angular momentum is. By choosing the lowest orbital momentum $L$, we see that $L=0$ for $J^P=1^+$ and $L=1$ for $J^P=0^+,2^+$. However, the dynamics will be different for $J^P=0^+, 1^+, 2^+$. This means that we can relate $\bar B^0_s \to Y(3940) \phi$ with $\bar B^0 \to Y(3940) \bar K^{*0}$, $\bar B^0_s \to Z(3930) \phi$ with $\bar B^0 \to Z(3930) \bar K^{*0}$, $\bar B^0_s \to X(4160) \phi$ with $\bar B^0 \to X(4160) \bar K^{*0}$ and $\bar B^0_s \to Y_P \phi$ with $\bar B^0 \to Y_P \bar K^{*0}$, but in addition we can relate $\bar B^0_s \to Z(3930) \phi$ with $\bar B^0_s \to X(4160) \phi$, and the same for $\bar B^0 \to Z(3930) \bar K^{*0}$ with $\bar B^0 \to X(4160) \bar K^{*0}$. Hence in this latter case we also have a $2^+$ state for both resonances and the only difference between them is the different coupling to $D^* \bar D^*$ and $D^*_s \bar D^*_s$, where the $Z(3930)$ couples mostly to $D^* \bar D^*$, while the $X(4160)$ couples mostly to $D^*_s \bar D^*_s$.

The partial decay width of these transitions is given by
\begin{equation}\label{eq:width}
\Gamma_{R_i}=\frac{1}{8\pi} \frac{1}{m^2_{\bar B_i^0}} \left| t_{\bar B_i^0 \to \phi (\bar K^{*0}) R_i} \right|^2 P^{2L+1}_{\phi (\bar K^{*0})},
\end{equation}
which allows us to obtain the following ratios, where the different unknown constants $V_P$, that summarize the production amplitude at tree level, cancel in the ratios:
\begin{eqnarray}
&&R_1 \equiv \frac{\Gamma_{\bar B^0_s \to \phi R^{J=0}}}{\Gamma_{\bar B^0 \to K^{*0} R^{J=0}}},~~~~
R_2 \equiv \frac{\Gamma_{\bar B^0_s \to \phi R^{J=1}}}{\Gamma_{\bar B^0 \to K^{*0} R^{J=1}}}, \nonumber \\
&&R_3 \equiv\frac{\Gamma_{\bar B^0_s \to \phi R^{J=2}_1}}{\Gamma_{\bar B^0 \to K^{*0} R^{J=2}_1}},~~~~
R_4 \equiv\frac{\Gamma_{\bar B^0_s \to \phi R^{J=2}_2}}{\Gamma_{\bar B^0 \to K^{*0} R^{J=2}_2}}, \nonumber
\end{eqnarray}
and
\begin{equation}
%\label{eq:ratio-3}
R_5 \equiv \frac{\Gamma_{\bar B^0_s \to \phi R^{J=2}_1}}{\Gamma_{\bar B^0_s \to \phi R^{J=2}_2}}, \nonumber
\end{equation}
where $R^{J=0}$, $R^{J=1}$, $R^{J=2}_1$ and $R^{J=2}_2$ are the $Y(3940)$, $Y_P$, $Z(3930)$ and $X(4160)$, respectively.
\section{Results}
The couplings $g_{MM', R}$ and the loop functions $G_{MM'}$ in Eq. (\ref{eq:amplituT}) are taken from ref. \cite{raquelxyz}, where the dimensional regularization was used to deal with the divergence of $G_{MM'}$, fixing the regularization scale $\mu =1000$ MeV and the subtraction constant $\alpha = -2.07$. However, we have taken advantage to make corrections to the work of ref.~\cite{raquelxyz} due to the findings of ref.~\cite{xiaoliang} concerning
  heavy quark spin symmetry. It was found there that a factor $m_{D^*}/m_{K^*}$ has to be implemented in the hidden gauge coupling
  $g=m_\rho/2 f_\pi$ in order to account for the $D^*\to D\pi$ decay. However, this factor should not be implemented in the
  Weinberg-Tomozawa terms (coming from exchange of vector mesons) because these terms automatically implement this factor in the
  vertices of vector exchange. However in the work of ref.~\cite{raquelxyz} the coupling needed for $D^*\to D\pi$ had been used also in the
  Weinberg-Tomozawa term. The bindings were obtained by fitting subtraction constants in the $G$ function. We also do the same now, but
   with the present reduced interaction the $G$ function becomes more negative. We use now $\mu=1000$ MeV and $\alpha=-2.19$.

In Eqs. (\ref{eq:ratio-1})-(\ref{eq:ratio-2}) we summarize the results that we obtain,
\begin{equation}\label{eq:ratio-1}
R_1=0.95,~~
R_2=0.96,~~
R_3=0.95,~~
R_4=0.83,~~
%R_5=1.02.
\end{equation}
and
\begin{equation}\label{eq:ratio-2}
R_5=0.84.
\end{equation}

As we can see, all the ratios are of the order of unity, however, the simplicity of these results should not be confused with a triviality. The ratios close to unity for the $\phi$ or $K^{*0}$ production are linked to the fact that the resonances are dynamically generated from $D^* \bar D^*$ and $D^*_s \bar D^*_s$, which are produced by the hadronization of the $c\bar c$ pair. The ratio for the $J^P=2^+$ is even more subtle since it is linked to the particular couplings of these resonances to $D^* \bar D^*$ and $D^*_s \bar D^*_s$, which are a consequence of the dynamics that generates these states. Actually, the ratios $R_1,~R_2,~R_3, ~R_4$ are based only on phase space and result from the elementary mechanisms of Fig \ref{Fig:1}. We would get the same ratios as far as the resonances are $c \bar c$ based. Hence, even of these ratios do not prove the molecular nature of the resonances, they already provide valuable information telling us that they are  $c \bar c$ based.

   The ratio $R_5$ provides more information since it involves two independent resonances and it is  not just a phase space ratio. If we take into account only phase space then $R_5\approx 4$ instead of the value $0.84$ that we obtain.

As for the absolute rates, we can establish an analogy to the $\bar B^0_s \to J/\psi f_0(980)$ decay, since the amplitudes of the quark level prior to hadronization are identical and both processes require hadronization of a $q\bar q$ pair into resonances. Hence, we estimate the branching ratio for the production of these states of the order of $10^{-4}$ \cite{pdg}. Another estimate can be done starting from $\bar B^0_s \to J/\psi \phi$, which has a branching ratio of $1.07 \times 10^{-3}$ and reducing this rate by one order of magnitude which is the reduction factor that is found for hadronization in ref. \cite{Bayar:2014qha}. In both cases we find branching fractions of the order of $10^{-4}$, which are an order of magnitude bigger than many rates of the order of $10^{-5}$ already catalogued in the PDG \cite{pdg}.

 Given the fact that the ratios $R_1,~R_2,~R_3, ~R_4$ obtained are not determining the molecular nature of the resonances, but only on the fact that they are $c \bar c $  based, we propose a complementary test in the next section.

\section{Complementary test of the molecular nature of the resonances}
 In this section we propose a test that is linked to the molecular nature of the resonances.
We study the decay $\bar B^0_s \to \phi D^* \bar D^*$ or $\bar B^0_s \to \phi D_s^* \bar D_s^*$ close to the  $D^* \bar D^*$ and $D_s^* \bar D_s^*$ thresholds.

\begin{figure*}[ht]
\begin{center}
\includegraphics[scale=0.5]{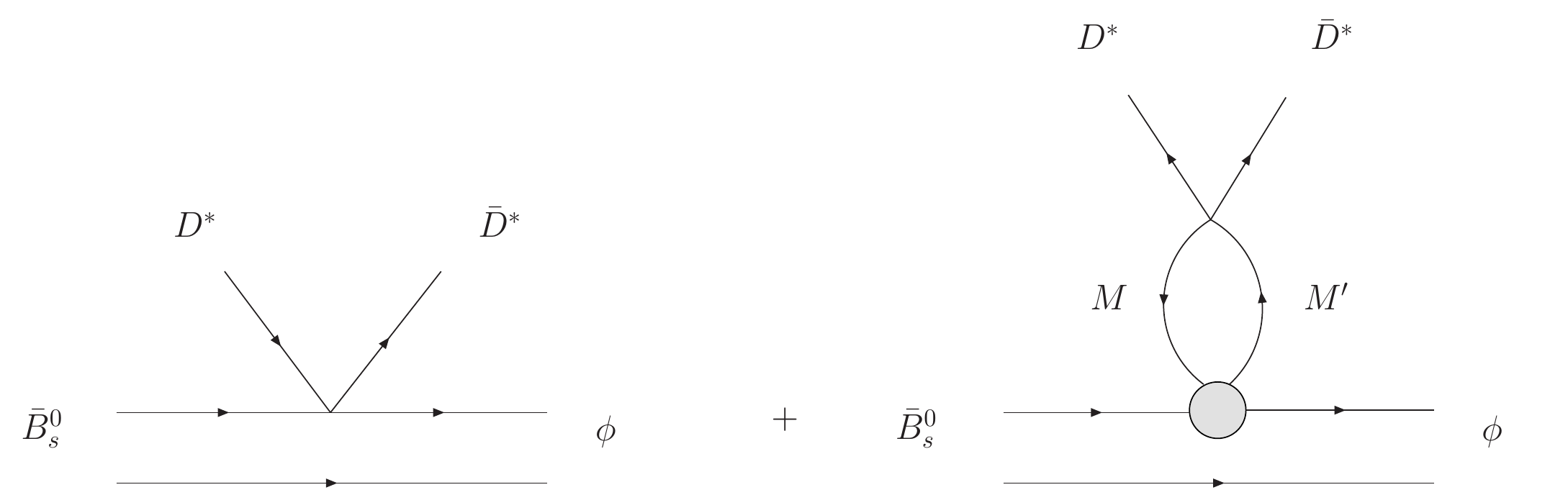}
\end{center}
\caption{Feynmann diagrams for the $D^*D^*$ production in $B^0_s$ decays.}
\label{new}
\end{figure*}

Let us now look to the process $\bar B^0_s \to \phi D^* \bar D^*$ depicted in Fig. \ref{new}.
%\lipsum[1-2]
%\lipsum[3-10]
The production matrix for this process will be given by
\begin{eqnarray}
t_{(\bar B^0_s \to\phi D^*\bar{D}^*)} =&& V_P (\sqrt{2}+\sqrt{2}\, G_{1}\,t_{(1\to 1)} + G_{2}\,t_{(2\to 1)}),\nonumber\\
\label{eq:tmat}
\end{eqnarray}
where $1$ and $2$ stands for the $D^*\bar{D}^*$ and $D^*_s\bar{D}^*_s$ channels respectively. The differential cross section for production will be given by \cite{weihong}
\begin{equation}
\frac{d\Gamma}{d M_{\mathrm{inv}}}=\frac{1}{32\,\pi^3 M^2_{\bar B^0_s }}p_\phi \tilde{p}_{D^*}|t_{(\bar B^0_s \to\phi D^*D^*)}|^2 p^{2 L}_\phi,
\label{eq:invd}
\end{equation}
where $p_{\phi}$ is the $\phi$ momentum in the $\bar B^0_s$  rest frame and $\tilde p_{D^*}$ the $D^*$ momentum in the $D^* \bar D^*$ rest frame. By comparing this equation with Eq. (\ref{eq:width}) for the coalescence production of the resonance in $\bar B^0_s \to \phi ~R$, we find
\begin{eqnarray}
R_\Gamma &&=\frac{M^3_{R}}{p_\phi \tilde{p}_{D^*}}\frac{1}{\Gamma_{R}}\frac{d \Gamma}{d M_{\mathrm{inv}}}\nonumber\\&&=\frac{M^3_{R}}{4 \pi^2}\frac{p_\phi^{2 L}(M_\mathrm{inv})}{p_\phi^{2 L+1}(M_{R})}\left| \frac{t_{(\bar B^0_s \to\phi D^*\bar{D}^*)}}{t_{(\bar{B}^0\to R\phi)}} \right|^2,
\label{eq:rg1}
\end{eqnarray}
where we have divided the ratio of widths by the phase space factor $p_{\phi} \tilde p_{D^*}$ and multiplied by $M_{R}^3$ to get a constant value at threshold and a dimensionless magnitude.
We would apply this method for the three resonances that couple strongly to $D^*\bar{D}^*$ (see Table \ref{Tab:XYZstate}). In the
case of the resonance $R_2$ with $J=2$ that couples mostly to the $D^*_s\bar{D}^*_s$ channel (see Table \ref{Tab:XYZstate}) we look
instead for the production of $D^*_s\bar{D}^*_s$.

\begin{figure*}[t]
\begin{center}
\includegraphics[scale=0.55]{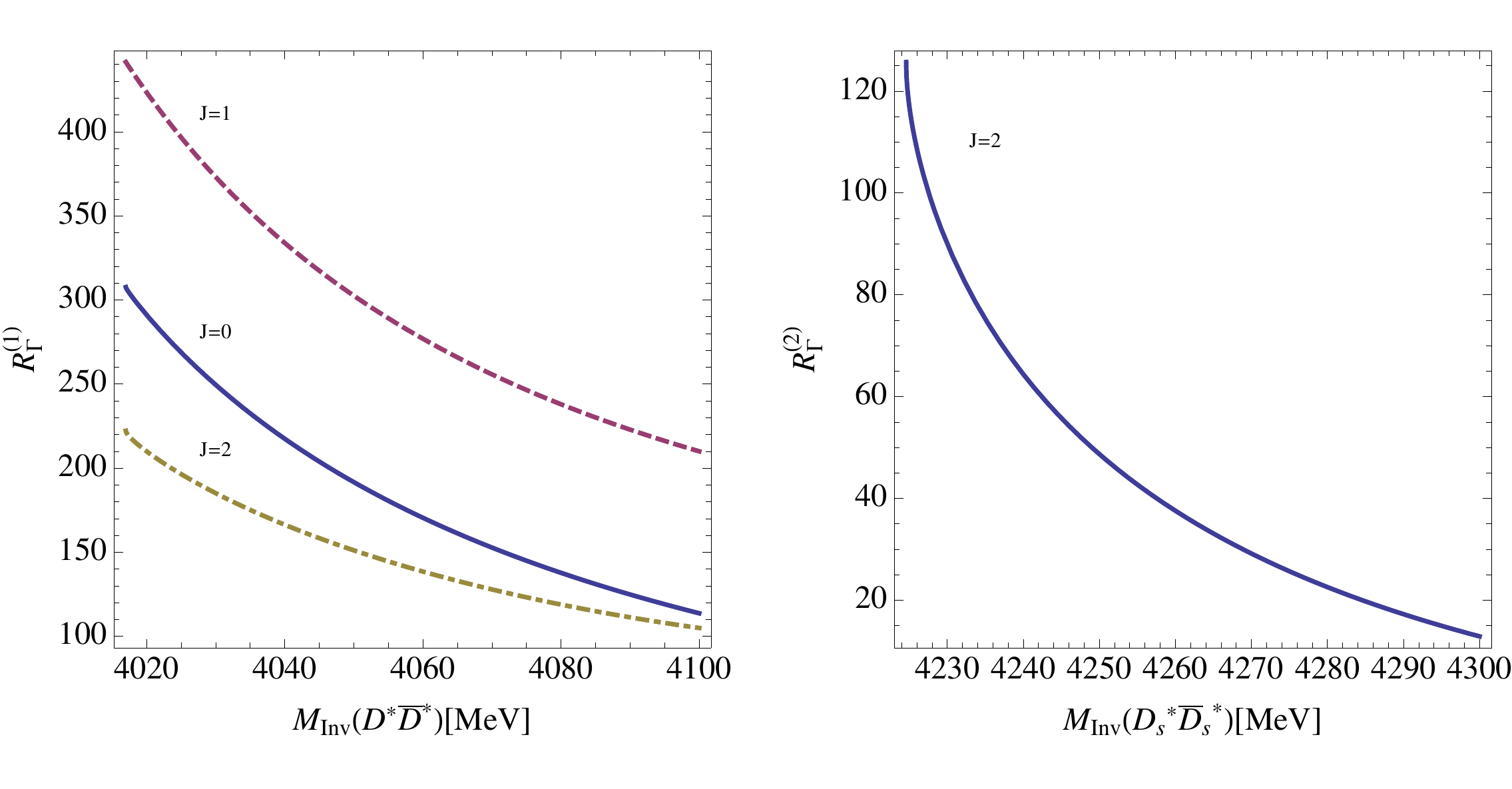}
\end{center}
\caption{Results of $R_{\Gamma}^{(1)}$ of Eq. (\ref{eq:rg1}) as a function of $M_{inv}(D^* \bar D^*)$ for the first three resonances of the Table \ref{Tab:XYZstate} (left) and $R^{(2)}_\Gamma$ as a function of $M_{inv}(D_s^* \bar D_s^*)$ (right) for the fourth resonance of the Table \ref{Tab:XYZstate}.}
\label{fig:rgam}
\end{figure*}

For the $D_s^* \bar D_s^*$ production we have
\begin{eqnarray}
t_{(\bar B^0_s \to\phi D^*_s\bar{D}^*_s)} =&& V_P (1+\sqrt{2}\,G_{1}\,t_{(1\to 2)}+ G_{2}\,t_{(2\to 2)}) ,\nonumber\\
\label{eq:tmat2}
\end{eqnarray}
and we use Eq. (\ref{eq:rg1}) but with $D^*_s\bar{D}^*_s$ instead of $D^*\bar{D}^*$ in the final state.
We have evaluated Eq. (\ref{eq:rg1}) using the scattering matrices obtained in ref.~\cite{raquelxyz} modified as discussed above, and then Eq. (\ref{eq:tmat}) and Eq. (\ref{eq:tmat2}). The results are shown in Fig. \ref{fig:rgam}.
%\lipsum[1-2]
%\lipsum[3-10]

We can see that the ratios are different for each case and have some structure. We observe that there is a fall down of the
differential cross sections as a function of energy, as it would correspond to the tail of a resonance below threshold. We should
also note that in the case of $D^*\bar{D}^*$, we have produced the $I=0$ combination. If instead, one component like $D^{*+}D^{*-}$ is observed,
the rate should be multiplied by $1/2$. In the case of $D^*_s\bar{D}^*_s$ there is a single component and the rate predicted is fine.

\begin{figure*}[ht]
\begin{center}
\includegraphics[scale=0.55]{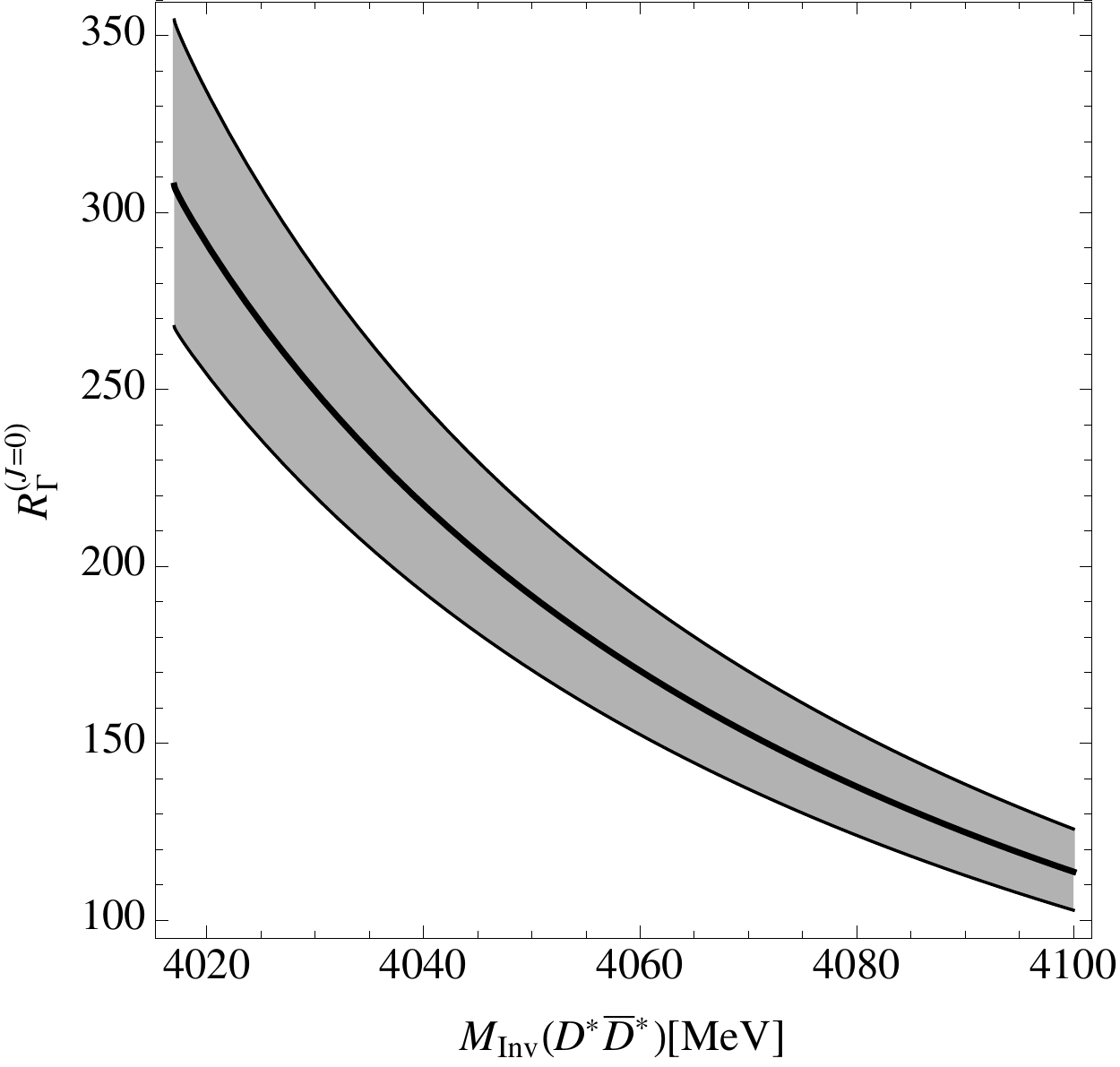}
\end{center}
\caption{Results of $R_{\Gamma}^{(1)}$
of Eq. (\ref{eq:rg1}) as a function of $M_{inv}(D^* \bar D^*)$ for spin$=0$.}
\label{fig:rgamj0}
\end{figure*}

In order to estimate uncertainties, we modify the model of ref.~\cite{raquelxyz} by changing a bit the value of $\alpha$ in the $G$ function expressions such that the values of the masses of the states change $5$ MeV up or down. The results are shown in Fig. \ref{fig:rgamj0}.

%\lipsum[1-2]
%\lipsum[3-10]
We find differences of  the order of  $15$\% in the $D^*\bar{D}^*$, $D^*_s\bar{D}^*_s$ mass distributions, which we accept as systematic uncertainties.

   Concerning the uncertainties of the ratios, since $R_1,~R_2,~R_3, ~R_4$ come from phase space, there are no changes for these ratios when using the second model. The ratio $R_5$ is modified and we find now $R_5=0.84\pm 0.02$.

\section{Conclusions}
  We have investigated the decays of $\bar B^0 \to \bar K^{*0} R$ and  $\bar B^0_s \to \phi R$ with  R being the $X(4160)$, $Y(3940)$, $Z(3930)$ and a predicted $J=1$ resonances. These decays have not been yet investigated. We estimate them to have branching ratios of the order of $10^{-4}$. We used a model in which these states are dynamically generated from the vector-vector interaction, most notably the $D^* \bar D^*$ and  $D^*_s \bar D^*_s$ states. Within this model we could predict five ratios for the production of these states, where unknown dynamics of the production processes cancels in the ratios. The procedure used a mechanism in which we have quark production at the elementary level, followed by hadronization of one final $q \bar q$ pair into two vectors and posterior final state interaction of this pair of vector mesons. This mechanism has been tested successfully in many other decays and given us confidence on the fairness of the predictions made. Some ratios only tell us that the resonance is build from $c \bar c$ with or without extra hadronization. In order to offer some extra test for the molecular composition of these states, we have evaluated the invariant mass distributions for the $\bar B^0_s \to \phi D^* \bar D^*$ and $\bar B^0_s \to \phi D_s^* \bar D_s^*$ close to threshold and have made predictions for these magnitudes relative to the width for the production of the resonances. The experimental investigation of these decay modes, and comparison with the predictions made, would shed light on the nature of these resonances and we can only encourage the implementation of such experiments.

\section*{Acknowledgments}
One of us, E. O., wishes to acknowledge support from the Chinese
Academy of Science (CAS) in the Program of Visiting Professorship
for Senior International Scientists.
This work is partly supported
by the Spanish Ministerio de Economia y Competitividad and European
FEDER funds under the contract number FIS2011-28853-C02-01 and
FIS2011-28853-C02-02, and the Generalitat Valenciana in the program
Prometeo II-2014/068. We acknowledge the support of the European
Community-Research Infrastructure Integrating Activity Study of
Strongly Interacting Matter (acronym HadronPhysics3, Grant Agreement
n. 283286) under the Seventh Framework Programme of EU. This work is
also partly supported by the National Natural Science Foundation of
China under Grant Nos. 11165005 and 11475227. The Project
is Sponsored by the Scientific Research Foundation for the Returned
Overseas Chinese Scholars, State Education Ministry and the american
National Science Foundation, PIF grant No. PHY 1415459.

%\clearpage

\bibliographystyle{plain}

\end{document}